\documentclass[letter,12pt]{article}
\usepackage{graphicx,amssymb,amsmath,epstopdf,color,hyperref}

\def\beq{\begin{equation}}
\def\eeq{\end{equation}}
\def\bea{\begin{eqnarray}}
\def\eea{\end{eqnarray}}

\newcommand{\gsim}{\lower.7ex\hbox{$\;\stackrel{\textstyle>}{\sim}\;$}}
\newcommand{\lsim}{\lower.7ex\hbox{$\;\stackrel{\textstyle<}{\sim}\;$}}

\def\be{\begin{equation}}
\def\ee{\end{equation}}
\def\bea{\begin{eqnarray}}
\def\eea{\end{eqnarray}}

\begin{document}

{~}

\vspace{0.07in}

\noindent
\begin{center}

{\bf\large Comments on Brane Recombination,\\ Finite Flux Vacua, and the Swampland}

\vspace{1.3cm}
{ Jason Kumar${}^a$, James D. Wells${}^b$}

{\it ${}^a$Department of Physics and Astronomy \\
University of Hawai'i, Honolulu, HI 96822 USA}

{\it ${}^b$Leinweber Center for Theoretical Physics \\
University of Michigan, Ann Arbor, MI 48109 USA} \\

\end{center}

\vspace{2cm}

\noindent
{\it Abstract:} The Swampland program relies heavily on the conjecture that there can only be a finite number of flux vacua (FFV conjecture). Stipulating this FFV conjecture and applying it to some older work in flux vacua construction, we show that within a patch of the landscape the  FFV conjecture makes predictions on the nonexistence of otherwise viable nonperturbative objects arising from brane recombination. Future gains in direct nonperturbative analysis could, therefore, not only test this prediction but also test
portions of the Swampland program itself. We also discuss implications of a weaker FFV conjecture on the counting of flux vacua which predicts positivity of the brane central charge if the EFT analysis is to be qualitatively trusted.

\vfill\eject

\section{The finite flux vacua conjecture}

Recently, there have been a variety of conjectures regarding the structure of
the Swampland, which can be thought of as the set of effective field theories that
cannot be realized as low-energy limits of a consistent theory of quantum
gravity~\cite{Vafa:2005ui} (see~\cite{Brennan:2017rbf,Palti:2019pca} for reviews).
Some recent work has suggested that consistent theories of quantum gravity
can only lead to low-energy effective field theories with a finite number of fine-tuned
parameters~\cite{Heckman:2019bzm}.
In the context of 4D, $N=1$ compactifications, a key premise to these analyses
is the claim that there are at most a finite number of flux vacua (see, for example,~\cite{Blumenhagen:2004xx,Douglas:2006xy,Acharya:2006zw}).

The claim of ``finite flux vacua" is somewhat vague in the literature, and it is worthwhile formulating a precise conjecture and investigating the consequences of it.  One might as well start with the strongest version that does not appear to be in conflict with any known result that is also often implied at times in the flux vacua~\cite{Douglas:2006xy} and Swampland literature. The ``finite flux vacua conjecture" ({FFV conjecture}), we'll call it, states that given only an upper bound on
the magnitude of
the vacuum energy, the number of 4D $N=1$ vacua is finite.  There is good circumstantial
evidence for the FFV conjecture. It has gone through non-trivial checks in the context of particular
string compactifications, but it has not been rigorously justified yet. Further non-trivial tests are needed.

In this letter, we focus on the question of how to further stress test the FFV conjecture. We show below that a key test is to determine whether particular nonperturbative stable objects exist in 4D $N=1$ theories, whose existence would contradict the FFV conjecture upon which the Swampland program so heavily relies. At present, the computational technology to carry out this test is not available, but one can show that the  FFV conjecture is in conflict with naive counting of flux vacua using standard rules of flux vacua construction based on effective theory analysis. Nevertheless, the  FFV conjecture, if correct, gives us a qualitative understanding of how flux vacua behave when the effective theory analysis starts to break down, which in principle can be tested when more reliable nonperturbative analysis techniques are developed.

In sec.~\ref{sec:weakFFV}, we point out that there is a weaker form of the FFV conjecture implied in the literature. This weaker conjecture also gives implications to the consideration of flux vacua.  One finds that the EFT analysis can be correct and valid qualitatively, including its prospects for forming nonperturbative stable objects from brane recombination; however, to be consistent in the limit of a large number of vacua,
the central charge must be positive, which adds a new rigorous and non-trivial constraint on the EFT description as the brane charges increase.

\section{Naive compatibility with  FFV conjecture}

Analyzing the entire flux vacua landscape is of course impossible, but it is tractable to analyze a patch of the landscape restricted by a few criteria. For example, for the rest of this discussion, our analysis will be entirely within what we call the Marchesano-Shiu patch (MS-patch) of the string landscape~\cite{Marchesano:2004yq,Marchesano:2004xz} which are vacua derived from type IIB string theories compactified to 4D $N=1$ on a Calabi-Yau orientifold of  $T^6/Z_2\times Z_2$ that produces SM-like massless spectra (i.e., SM states plus arbitrary additional hidden-sector states or vector-like states\footnote{Note that although the MS-patch was formulated for the purpose of producing the Standard Model in
the low-energy limit, the presence of an SM-like spectrum is not necessary for the purpose of studying the FFV conjecture.}).
The specificity of this construction in no way limits our ability to devise a non-trivial test of the  FFV conjecture, since it is precisely within these MS-vacua that one can so clearly see a well-articulated potential conflict.  The finiteness of
flux vacua for the case of an orientifolded $T^6/Z_2\times Z_2$ was considered in~\cite{Blumenhagen:2004xx,Douglas:2006xy}.

In this section, we first  review the standard argument for why there are a finite number of flux vacua predicted using standard flux vacua construction techniques within the MS-patch. We show that intuitive results from an application of the rules of construction will be in concert with the  FFV conjecture.
Having gone through the standard argument, we will then point out a subtlety~\cite{Kumar:2006yg} which will impact the FFV conjecture.

The argument begins with a realization that a string compactification on an orientifolded $T^6/Z_2\times Z_2$
will yield a 4D $N=1$ theory with negative  RR-charges associated with the background provided by the orientifold planes.   In a consistent compactification, the space-filling charges must be cancelled, and the cancellation of RR-tadpoles {associated with the negative charges} requires the addition of positive charges, which can be supplied from additional $D$-branes (which preserve the same supersymmetry as the background) or from RR fluxes that wrap the compactified dimensions.  But the number of flux vacua scales with the RR flux since the amount of flux that is turned on is related to the number of solutions for the supersymmetric equations of motion of the complex structure moduli~\cite{Denef:2004ze}. The K\"ahler moduli are then fixed by a variety of nonperturbative contributions to the superpotential.

As mentioned above, the claim for the finiteness of the number of flux vacua originates from the fact that the contribution to space-filling $D$-brane charges from the orientifold planes is negative and finite.  If the only other space-filling charges arise from fluxes, then the amount of flux would also be finite, leading to a finite number of solutions for the complex structure moduli.  $D$-branes can also contribute space-filling charges, but this can only alleviate the problem if the charges are negative.  Anti-$D$-branes would contribute negative charge, but would also explicitly break supersymmetry.  Magnetized $D$-branes can contribute some negative charges, but necessarily also contribute positive charges.  It has thus been shown that the required cancellation between positive and negative contributions to the space-filling charges cannot arise from a set of branes that all preserve the same supersymmetry as the orientifold background at any point in the K\"ahler moduli space (see, for example,~\cite{Blumenhagen:2004xx,Douglas:2006xy}).
Thus, one is necessarily limited to a finite amount of flux that cancels the finite negative $D$-brane charges from the orientifold planes, and with finite flux comes a finite number of vacua.

\section{Naive incompatibility with  FFV conjecture}

There is a shaky premise used in the last section that enabled the proof of the compatibility of the  FFV conjecture with the counting of flux vacua within the MS-patch. Namely, we assumed that all D-branes introduced directly preserve the same supersymmetry. As pointed out in~\cite{Kumar:2006yg}, it may  not be necessary for all D-branes to preserve the same supersymmetry at one point in moduli space in order for the theory to have a supersymmetric vacuum.  Instead, the branes can deform via the condensation of open strings, leading to a     supersymmetric minimum.  In that case,  brane recombination can result in supersymmetric solutions in which a much more negative D-brane charge from brane bound states is canceled by a much larger RR-flux, yielding many more flux vacua~\cite{Kumar:2006yg}.  In fact, naive application of the rules of engagement suggests an unlimited amount of negative D-brane charge, cancelled by the equally unlimited RR-flux, resulting in an unlimited number of flux vacua within the MS-patch. If true, this would be in unambiguous conflict with the  FFV conjecture which says there must be a finite number of flux vacua not only within the MS-patch but summing over all patches of the landscape.

Let us discuss in more detail the origin and characteristics of these putative bound states from brane recombination. One can infer the possibility of a supersymmetry preserving bound state of branes by considering the low-energy effective field theory describing the light states.  At a point in K\"ahler moduli space where a brane preserves the same $N=1$ SUSY as the background, the light open strings are described by a supersymmetric gauge theory with no non-vanishing Fayet-Illiopoulos (FI) terms.  If one moves slightly away from this point in moduli space, then generically the brane will no longer preserve the same supersymmetry as the background; this is reflected in the low-energy effective field theory by the turning on of an FI-term, yielding a $D$-term potential of the general form~\cite{Berkooz:1996km,Kachru:1999vj}
\bea
V_{D} &=& \frac{1}{2g^2} \left(\sum q_i |\phi_i|^2 + \xi \right)^2 ,
\eea
where the scalars $\phi_i$ represent chiral open strings stretching between branes at their intersection points.
If none of these scalars have non-vanishing vacuum expectation values, then this $D$-term will be non-vanishing.
But if there exist charged scalars with appropriate charges, then they will be tachyonic and condense, allowing the $D$-terms to relax again to zero~\cite{Kachru:1999vj}.
The condensation of these scalars amounts to the process of brane recombination, in which the branes deform near their
intersection points to form supersymmetric bound states.

The question of whether or not a set of branes will recombine to form a SUSY bound state revolves around the signs of the FI-terms and the signs of the charged fields.  For any field with charge $q$ under a $U(1)$ gauge group, the FI-term $\xi$ will give a contribution to the squared mass of $q \xi$; this contribution is tachyonic if $q$ and $\xi$ have opposite sign.  If there are enough tachyons for all of the $D$-terms to relax to zero, then there is no obstruction to the formation of a SUSY bound state.

Indeed, if there are $N$ stacks of $D$-branes, all with small FI-terms, then one would generically expect the presence of ${\cal O}(N^2)$ chiral scalars arising from strings stretching between these branes at their points of intersection.  Given that the charges are determined by the signs of the topological intersection numbers of the branes, one would generically expect to find enough scalars with the correct charge sign assignments for all $N$ of the $D$-terms to relax to zero.

To illustrate these points in some more detail, we show an explicit construction that naively gives an infinite number of flux vacua within the MS-patch, and then discuss how the analysis is likely to break down at large fluxes, where one could then suppose that a more complete analysis brings the system into agreement with the  FFV conjecture. Again, the MS-patch considers the case of Type IIB string theory compactified on an orientifolded $T^6 / Z_2 \times Z_2$ yielding SM-like vacua. We follow the notation of~\cite{Cascales:2003zp}, and utilize the method of identification of large-flux vacua present in~\cite{Kumar:2005hf}.

This orientifold generates 64 $O3$-planes and 12 $O7$-planes, and preserves
$N=1$ supersymmetry. There are 3 K\"ahler moduli, and 51 complex structure moduli (48 of which lie at fixed points
of elements of the orbifold group). We will consider brane stacks with wrapping numbers $(n_i , m_i)$ given by~\cite{Kumar:2006yg}
\bea
N_q=2 &:& \qquad \left(-(x-1)^2,1 \right)~ \left(-(x-1)^2,1 \right)~ \left(-(x-1)^2,1 \right) ,
\nonumber\\
N_r =2 &:& \qquad (1,-1)~(1,x)~(1,x) ,
\nonumber\\
N_s =2 &:& \qquad (1,x)~(1,x)~(1,-1),
\nonumber\\
N_t =2 &:& \qquad (1,x)~(1,-1)~(1,x) ,
\eea
where the $N_i$ are the number of branes in each stack, $x$ is an integer,
and the $(n_i, m_i)$ are co-prime integers, with $m_i$ being the number of times the brane
wraps the $i$th torus, and $n_i$ being the quantized magnetic flux through the $i$th torus.
The total D-brane charges of these stacks are~\cite{Kumar:2006yg}
\bea
\left\{Q_{D3}, \overrightarrow{Q}_{D7_i} \right\} &=& \left\{2[-(x-1)^6+3], 2, 2, 2 \right\} ,
\eea
where $Q_{D7_i}$ is the charge of D7-branes wrapping every torus except the $i$th.
The total D-brane charges required from all of the branes and the fluxes in order to satisfy the tadpole constraints are
\bea
\left\{Q_{D3}, \overrightarrow{Q}_{D7_i} \right\} &=& \left\{16,16,16,16 \right\} ,
\eea
implying that one can add to the above brane stacks $D7$-branes, $D3$-branes, and an amount of flux (which contributes D3-brane charge)
that grows with $x$ in order to satisfy
all tadpole constraints.  It only remains to be seen if these magnetized branes can form a supersymmetric bound state.

We list the multiplicities of the chiral open strings:
\bea
(x^2-2x)(x^3-2x^2+x+1)^2 &\rightarrow& (1,-1,0,0) , (1,0,-1,0) , (1,0,0,-1) ,
\nonumber\\
2x(x-1)^2 &\rightarrow& (0,-1,0,-1) , (0,-1,-1,-0) , (0,0,-1,-1) ,
\nonumber\\
(x^2-2x+2)(x^3 - 2x^2 +x-1)^2 &\rightarrow& (-1,-1,0,0), (-1,0,-1,0), (-1,0,0,-1),
\nonumber\\
2x^2 - 4x +2 &\rightarrow& (0,2,0,0),(0,0,2,0),(0,0,0,2) ,
\nonumber\\
4(x-1)^6 - 2(3(x-1)^4-1) &\rightarrow& (2,0,0,0),
\eea
where the expressions on the left are multiplicities, and the numbers on the right are the
charges $(Q_q, Q_r, Q_s, Q_t)$.
For $x$ reasonably large, all of these multiplicities are positive (a negative multiplicity would count the
number of charge conjugate fields).

The FI-terms, at leading order, are given by
\bea
\xi_a \sim \sum_{i=1}^3 \tan^{-1} (m_i^a {\cal A}_i , n_i) ~\mod 2\pi,
\eea
where the ${\cal A}_i$ are the real K\"ahler moduli which determine the volume of the tori, and where
$\alpha = \tan^{-1} (y,x) $ for $\sin \alpha = y/\sqrt{x^2 + y^2}$, $\cos \alpha = x/\sqrt{x^2 + y^2}$
(see, for example,~\cite{Dudas:2006vc}).
But this is the leading order calculation, and can only be trusted when the FI-terms are small.

The $D$-term potential is given by
\bea
V_{D_a} &=& \frac{1}{2g^2} \left(\sum q_i |\phi_i|^2 + \xi_a \right)^2 .
\label{eqn:DtermPot}
\eea
We see that we can cancel the $D$-term potential no matter what the FI-terms are. This realization comes by recognizing that
there are scalars which have negative charge under two $U(1)$s (for any choice of the
two groups); by turning them on, we can be sure to add an arbitrarily large negative
contribution to each FI-term which then can counter the $\xi_a$ values no matter how large they might be.
In other words, for each $D$-term, we can be sure that the sum of the FI-term and the
contribution from these scalar is non-positive.  But there are also scalars with
charge +2 under each group.  We can tune them independently to give a positive contribution
inside the square which cancels each $D$-term.

We still need to worry if a non-zero $F$-term will arise from turning on
these scalars.  But one simple way of addressing this is to note that the only
gauge-invariant superpotential terms which we can write involve at least three of the
scalars; such a term can only introduce a non-vanishing $F$-term if at least all but one
of the scalars get a vev.  But the three scalars will live at three different
intersections of the brane stacks, and the coefficient of the superpotential term
will be exponentially suppressed by the volume of the triangle formed by these
intersection points (in string units).  As long as we make sure that the scalars which
we turn on are far separated in the compact dimensions, the $F$-terms will be exponentially
suppressed.

Note that these solutions would be valid, supersymmetric solutions
in the effective field theory, and the flux can increase to arbitrarily high values with $x$ which then
indicates an arbitrarily large number of flux vacua, inconsistent with the  FFV conjecture.
One might worry that the exponentially-suppressed Yukawa couplings might lead to solutions
for which $N=1$ supersymmetry was instead broken at an exponentially suppressed scale, but again we
would be led to an arbitrarily large number of such solutions, in conflict with the FFV conjecture.

\section{Nonperturbative predictions from  FFV conjecture}

So far, we have reviewed the standard argument that the construction of flux vacua within the MS-patch is consistent with the FFV conjecture, and then we reviewed above a finding from~\cite{Kumar:2006yg} that if supersymmetric brane recombination is allowed, and there appears to be no obvious obstruction against doing that, there can be an infinite number of flux vacua within the MS-patch which is incompatible with the  FFV conjecture. In this section, we analyze this second claim more critically to determine if all the dynamics that we have identified as needed to make the ever-growing large flux are justified.

In fact, there is a significant subtlety that we have thus far ignored that comes to the core of the issue of whether these nonperturbative objects of brane recombination are allowed. We have assumed that all of the light degrees of freedom  may be determined by simply counting the topological intersection numbers of the undeformed branes and their orientifold images, which determine the lightest degrees of freedom living at the intersections of the branes.  This is guaranteed to be true if the undeformed branes are supersymmetric, or nearly so, which amounts to saying that the FI-terms must be small in string units.  However, for the type of brane stacks that are required here --- those that yield large negative $D3$-brane charge --- there need not be any place in moduli space where all of the FI-terms are small.

As one moves far away from the point in moduli space where a single brane preserves the same supersymmetry as the background, and the FI-term becomes of the same order as the string scale,
then the original effective field theory description is no longer valid. The effective field theory can break down in several ways, including the fact that the original light degrees of freedom could become heavy, while new degrees of freedom become light.  Indeed, the Swampland Distance Conjecture~\cite{Ooguri:2006in} is essentially the claim that this will be the case.
Thus, the form of the $D$-term
potential may be significantly corrected from that given in eq.~\ref{eqn:DtermPot}, as one moves a significant distance in moduli space.

Despite the detailed calculations necessarily breaking down with large flux, the larger qualitative question which remains is whether these putative supersymmetric bound states do in fact exist.  To address this question, one would need a formalism for describing the stable objects (and their degrees of freedom) as one moves arbitrarily far in moduli space.  Although $\Pi$-stability is a step in this direction for the $N=2$
case~\cite{Douglas:2000ah}, there is no such framework available yet for the case of $N=1$.

Our view, instead, is that the  FFV conjecture corresponds to a claim that such supersymmetric bound states {\it should not} exist.  Note, this does not amount to merely the claim that for a suitably large excursion of the moduli the FI-terms receive large, uncomputable corrections.  Although this would be expected, it would not resolve the problem; as we have shown, generically, we would expect enough scalars to become tachyonic and allow the $D$-terms to relax to zero for {\it any} choice of the FI-terms.  The problem is that the low-energy theory itself may be different, and we would have no way of counting the degrees of freedom to determine if there exist enough tachyons to relax the potential to zero.
Indeed, we would not necessarily know how many $D$-terms there are, as the rank of the gauge group may also change if one moves a large
distance in moduli space.
The claim here is that  FFV conjecture insists that these tachyons are eliminated and the number of flux vacua remains
large but finite.
Other recent work has also proposed non-trivial constraints on supersymmetric nonperturbative objects in theories
which do not belong to the Swampland~\cite{Abe:2020jgf}.

\section{Weak-FFV conjecture and EFT analysis}
\label{sec:weakFFV}

We have discussed at length a rather strong version of a FFV conjecture, but there are weaker versions of the conjecture that could be compatible with the EFT analysis as long as an additional constraint on the central charge is imposed. We can define the weak-FFV conjecture to be one that  states that given an upper bound on  the vacuum energy, an upper bound on the compactification volume, and a lower bound on the KK mass scale, the number of 4D vacua is finite. This weaker version was all that was needed to agree to some flux vacua analyses in the literature (e.g., see~\cite{Acharya:2006zw}).

Let us suppose that the specific $T^6/Z_2\times Z_2$ set-up that we discussed above is required to adhere only to this weak-FFV but not the stronger version. What might be the consequences? To answer this, we must recognize that as the charges increase dramatically to achieve an increasing number of flux vacua, the brane central charge
\bea
Z &=& Q_{D3} + Q_{D7_1} {\cal A}_2 {\cal A}_3 + Q_{D7_2} {\cal A}_3 {\cal A}_1 + Q_{D7_3} {\cal A}_1 {\cal A}_2 ,
\eea
 is prone to go negative rapidly for fixed compactification volume since $Q_{D3}\to -x^6\to -\infty$ while the $Q_{D7_i}$ stay fixed. With no requirement on the central charge $Z$, the number of flux vacua increases in violation of even the weak-FFV conjecture, unless we posit that the EFT analysis breaks down as we did in the stronger FFV conjecture above. For the EFT analysis to still be valid --- to still qualitatively describe the degrees of freedom and predict finite vacua --- one must impose a positivity constraint on the central charge $Z>0$. In this case, the central charge can only be positive when ${\cal A}_i$ are large. For example, for the case of ${\cal A}_i={\cal A}$, one requires ${\cal A} > x^3$ to enable $Z>0$ required for self-consistency.  Thus, it is clear that the recombination of a set of many branes, with a net $D3$-brane charge which is arbitrarily negative, forces us  to an arbitrarily large volume (decompactification limit), which at some point becomes incompatible with any finite parameters applied to the weak-FFV. Although the decompactification limit cannot be  analyzed in the language of the low-energy effective field theory once some limit is crossed, the qualitative compatibility with the weak-FFV has been shown within the EFT framework since for any fixed compactification volume there are indeed only a finite number of flux vacua allowed by the straightforward implementation of the EFT analysis presented above.

But it is worth noting that the conjecture for a finite number of fine-tunings seems to derive only from the strong FFV conjecture, not
the weak FFV conjecture.  As the compactification volume becomes increasingly large, the KK scale becomes smaller and smaller, but for
any finite compactification volume, one is left with a 4D theory in deep IR.  If only  the weak FFV conjecture holds, then one would
find that 4D $N=1$ theories could have an arbitrarily large number of fine-tuned parameters, realized in theories with increasingly small
(though finite) KK scales.  Sufficiently deep in the IR, one could obtain as many fine-tuned parameters as one desired.  In particular,
if only the weak FFV holds, then given any putative limit on the number of fine-tuned parameters of the low-energy theory, there exists a
scale such that, in the 4D effective field theory
defined below this scale, the number of fine-tuned parameters exceeds the putative limit.  We thus see a connection between the strength
of the FFV conjecture, and the precise formulation of the conjectured limit on the number of fine-tuned parameters.

\noindent
{\it A Few Subtle Issues}

Thus far, we have ignored a few additional subtle issues on the grounds that they largely do not affect our main point.  We
now discuss a few of these points here, to emphasize that these issues illustrate our main point, which is that
the FFV conjecture requires consistency conditions for nonperturbative states that are not
obvious from the point of view of the low-energy effective field theory.

First, it is worth noting that the $D$-term potential does not have true FI-terms, since they depend on the K\"ahler
moduli (see also~\cite{Komargodski:2009pc}).  But this does not affect our argument.
Though in principle one should fix all moduli simultaneously, we consider here for simplicity
the scenario in which the K\"ahler moduli are stabilized at a higher scale (see,
for example,~\cite{Binetruy:2004hh,Diaconescu:2007ah,Sinha:2007rg}).
But this assumption is not critical to our point.  Although consistency with the FFV conjecture would be maintained if there was an obstruction to the effective FI-terms obtaining the appropriate signs when the
open-string moduli solved the $D$-term equations, this would only confirm our point, which is that the FFV conjecture requires
a consistency condition for nonperturbative states which is non-trivial from the effective field theory point of view.

As a second point, we have utilized the fact that
the Yukawa couplings can be exponentially suppressed.
But these couplings could still become large in the IR, through effects
akin to dimensional transmutation.  But nothing requires this from the effective field theory
point of view; a requirement that such couplings run to a large value
and generate a non-vanishing
$F$-term would be another example of a consistency condition for nonperturbative states which
is non-obvious from the EFT point of view, but which is required for the FFV conjecture.

We have suggested a criterion that might allow consistency with the weak FFV conjecture, namely, that the
central charge for bound states preserving the same supersymmetry as the background be positive.  Although
the lowest-order expression for the central charge would indeed require an increasingly large volume in order
for states with increasingly negative $D$-brane charge to still have positive central charge, this expression
will get large corrections at small volume.
These corrections in general may involve more than the volume moduli, including Wilson line moduli,
for example.
As such, the criterion we suggested may indeed fail to rescue the
weak FFV, as these problematic states might have positive central charge at small volume.  In this case, some corrected
version of this criterion would be required.  Our goal was not to present definitively the criterion that must
be satisfied for a nonperturbative state to preserve the same supersymmetry as the background, as this criterion is not known.
Rather, our point was to suggest the type of criterion that could provide consistency with the weak FFV.

We do not intend to suggest that our discussion of these issues is
exhaustive.  There are many issues related to the counting of flux
vacua that we have not addressed.  For example, though it is often
believed that the number of Calabi-Yau three-folds is finite, this
has not been proven.  Moreover, we have considered a particular
choice of orientifold projection, yielding orientifold planes with
particular charges, but there are others.  We have addressed a particular
construction of flux vacua, and have suggested criteria that could provide
consistency with the weak FFV.  It is not obvious how this type of criterion
should be generalized in order to understand string vacua which do not arise from this
type of construction.  For example, there are a variety of stringy vacua for
which the low-energy analysis of this type is inapplicable.  If these vacua
are also consistent with forms of the FFV conjecture, it is not clear how they
would be connected to the central charge criteria we have discussed here. Nevertheless, the patch of the landscape that we have discussed presents unique opportunities to test landscape ideas through nonperturbative analysis, which is why we focused on it.

\section{Summary}

In summary, we have seen in this analysis a possibly interesting interplay between a detailed formulation of the conjecture(s) of a finite number of 4D $N=1$ flux vacua
and the structure of nonperturbative supersymmetric objects in string theory.  This leads to the possibility that a better
understanding of the viability and stability of supersymmetric brane bound states could lead to concrete tests of some Swampland conjectures.

\bigskip

\noindent
{\bf Acknowledgements}
We are grateful to Emilian Dudas and Jonathan Heckman for helpful discussions.
The work of JK is supported in part by DOE grant DE-SC0010504, and JW by DOE grant DE-SC0007859.


\end{document}